\documentclass[aps,prl,10pt,twocolumn,tightenlines,amsmath,amssymb,superscriptaddress,floatfix]{revtex4-1}

\usepackage[english]{babel}
\usepackage[utf8]{inputenc}
\usepackage{amsmath,amssymb,amstext}
\usepackage{mathtools}
\usepackage{graphicx}
\usepackage{mdframed}
\usepackage{dsfont}
\usepackage[final,colorlinks=true,linkcolor=blue,citecolor=blue]{hyperref}    
\usepackage{epstopdf}

\newcommand{\ket}[1]{\vert#1\rangle}
\newcommand{\bra}[1]{\langle#1\vert}

\newcommand*\rfrac[2]{{}^{#1}\!/_{#2}}

\newcommand*{\e}[1]{\operatorname{e}^{#1}}
\newcommand*{\abs}[1]{\lvert #1 \rvert}
\newcommand{\ii}{\mathrm{i}}
\renewcommand*\d[1]{d#1\,}
\newcommand*\conv{\mathbin{*}}
\let\Re\relax
\DeclareMathOperator{\Re}{Re}

\newcommand{\be}{\begin{equation}}
\newcommand{\ee}{\end{equation}}
\newcommand{\bef}{\begin{figure}[!t]\centering}
\newcommand{\eef}{\end{figure}}
\newcommand{\beff}{\begin{figure*}[!t]\centering}
\newcommand{\eeff}{\end{figure*}}
\newcommand{\bs}{\begin{split}}
\newcommand{\es}{\end{split}}
\newcommand{\bt}{\begin{center}
\begin{tabular}}
\newcommand{\et}{\end{tabular}
\end{center}}

\begin{document}

\title{
Hectometer revivals of quantum interference
}

\author{Markus Rambach}
\email{mrks.rambach@gmail.com}
\affiliation{ARC Centre for Engineered Quantum Systems, School of Mathematics and Physics, University of Queensland, Australia.} 

\author{W. Y. Sarah Lau}
\affiliation{ARC Centre for Engineered Quantum Systems, School of Mathematics and Physics, University of Queensland, Australia.} 

\author{Simon Laibacher}
\affiliation{Institut f\"{u}r Quantenphysik and Center for Integrated Quantum Science and Technology (IQ\textsuperscript{ST}), Universit\"{a}t Ulm, Germany.} 

\author{Vincenzo Tamma}
\affiliation{Faculty of Science, SEES and Institute of Cosmology \& Gravitation, University of Portsmouth, UK.} 
\affiliation{Institut f\"{u}r Quantenphysik and Center for Integrated Quantum Science and Technology (IQ\textsuperscript{ST}), Universit\"{a}t Ulm, Germany.} 

\author{Andrew G. White}
\affiliation{ARC Centre for Engineered Quantum Systems, School of Mathematics and Physics, University of Queensland, Australia.} 

\author{Till J. Weinhold}
\affiliation{ARC Centre for Engineered Quantum Systems, School of Mathematics and Physics, University of Queensland, Australia.} 

\begin{abstract}
Cavity-enhanced single photon sources exhibit mode-locked biphoton states with comb-like
correlation functions.
Our ultrabright source additionally emits either single photon pairs or two-photon NOON states, dividing the output into an even and an odd comb respectively.
With even-comb photons we demonstrate revivals of the typical non-classical Hong-Ou-Mandel interference up to the
84th dip, corresponding to a path length difference exceeding 100~m. 
With odd-comb photons we observe single photon interference fringes modulated over twice the displacement range of the Hong-Ou-Mandel interference.
\end{abstract}

\maketitle

The Hong-Ou-Mandel (HOM) effect~\cite{Ou1987, Hong1987,Shih1988}---where photons in separate spatial modes coalesce after interfering at a beam splitter---is the most famous signature of non-classical interference. It varies directly with the indistinguishability of the interacting light fields in all degrees of freedom. This effect is inherently quantum and foundational in many quantum applications including photonic entangling gates~\cite{Knill2001,OBrien2003, Langford2005, Kok2007, OBrien2009}, measurement processes~\cite{Pan1998, Kwiat1998, Pryde2004} and boson sampling~\cite{Broome2013, Spring2012, Tillmann2013,Crespi2013}, and can also be used to measure the temporal width of the photonic wavepacket~\cite{Hong1987} or perform (sub-)femtosecond spectroscopy~\cite{Chen2015,Kalashnikov2017,Lyons2018}.  
The result of HOM interference is a low-order NOON state~\cite{Lee2002, Dowling1998}, which is of great interest in metrology as it has both phase super-resolution and phase super-sensitivity~\cite{Resch2007}.

The sensitivity of HOM interference to distinguishability in all degrees of freedom makes it a useful sensor to detect phase drifts or displacements, e.g. see \cite{Schwarz2011}. 
Most commonly a decrease in the coincidence rate --- a HOM dip --- is observed by varying the arrival time of the photons or the path lengths traversed by the light fields. Traditional single photon sources based on atoms, quantum dots or parametric down conversion will exhibit a single dip with a width representative of the coherence time of the two-photon state, usually on the order of pico- to femto-seconds.

Lu, Campbell and Ou~\cite{Lu2003} showed that placing a cavity before the interfering beam splitter leads to \textit{revivals} of the HOM interference, spaced by the round trip time of the cavity. The photons are then generated in a mode-locked entangled state with selected frequencies and a distinct temporal profile, causing a revival for every possible temporal output mode of the cavity. Assuming the cavity contains a single excitation, the intensity of the output field is exponentially decaying and thus results in a diminished visibility of the interference for longer delay times. A cavity around a single photon source produces the same effect and enhances the photon rate. To date, this effect has only been demonstrated over the range of several centimetres and a maximum of 9 revivals~\cite{Xie2015}.

Here, we present the first source that produces non-classical interference for ${>}100$~metres path difference between photons, and measure HOM interference  out to the $84^{\mathrm{th}}$ revival.

\emph{Biphoton frequency comb.} --- We produce frequency-entangled single photon pairs at 795~nm by cavity-enhanced spontaneous parametric down conversion (SPDC), achieving a spectral brightness of $(4.4~{\pm}~0.4){\times}10^{3}~\textrm{photon~pairs} / \textrm{(s~mW~MHz)}$~\cite{Rambach2017}. The pump light at 397.5~nm is generated inside a separate cavity with a linewidth $\sim$3~MHz.  We then type-II quasi-phase match in a periodically poled potassium titanyl phosphate (ppKTP) crystal to obtain the photon pairs.  Birefringence is compensated by the flip trick~\cite{Rambach2016}: a half-wave plate (HWP) at $45^\circ$ inside the bow-tie cavity flips the polarisation of each down-converted photon once per \emph{physical} round trip of temporal length $T_p$. 
Two physical round trips, one in each polarization, become one \emph{effective} round trip, referenced henceforth simply as round trip with $T \equiv 2T_p$. 

The biphoton frequency comb spans the 100~GHz full-width half maximum phase-matching bandwidth of the crystal and thus contains approximately 800 frequency modes, spaced by 120.8~MHz --- the free spectral range (FSR) of the cavity. As signal and idler photons satisfy the resonance condition simultaneously, the linewidth of the frequency modes are smaller than the width of the cavity resonance~\cite{Scholz2009b}, which are $429 \pm 10$~kHz for the modes and $666 \pm 15$~kHz for the cavity respectively~\cite{Rambach2016,Rambach2017a}. This corresponds to a coherence time for the heralded single photons of $\tau_{coh}=740 \pm 20$~ns~\cite{Rambach2017}, enabling the observation of quantum effects between two photons with temporal delays up to this magnitude.

We implement temporal delays for our experiments through a combination of different methods, as shown in Fig.~\ref{fig:delay}. \emph{Coarse} delays select which HOM dip is observed, \emph{intermediate} delays allow scanning over individual dips, and \emph{fine} delays enable observation of sub-wavelength scale features.  Coarse delays are introduced by the addition of optical fibres matched to multiples of the physical cavity round trip time of $T_p=T/2=4.14$~ns ($\sim$metres). Intermediate delays are on the order of picoseconds ($\sim$millimetres), achieved in free-space with two mirrors on a motorized translation stage before the photons are coupled into fibre. Fine delays on the order of femtoseconds ($\sim$nm) are achieved using a piezo-mounted mirror. 

\bef
\includegraphics[width=\linewidth]{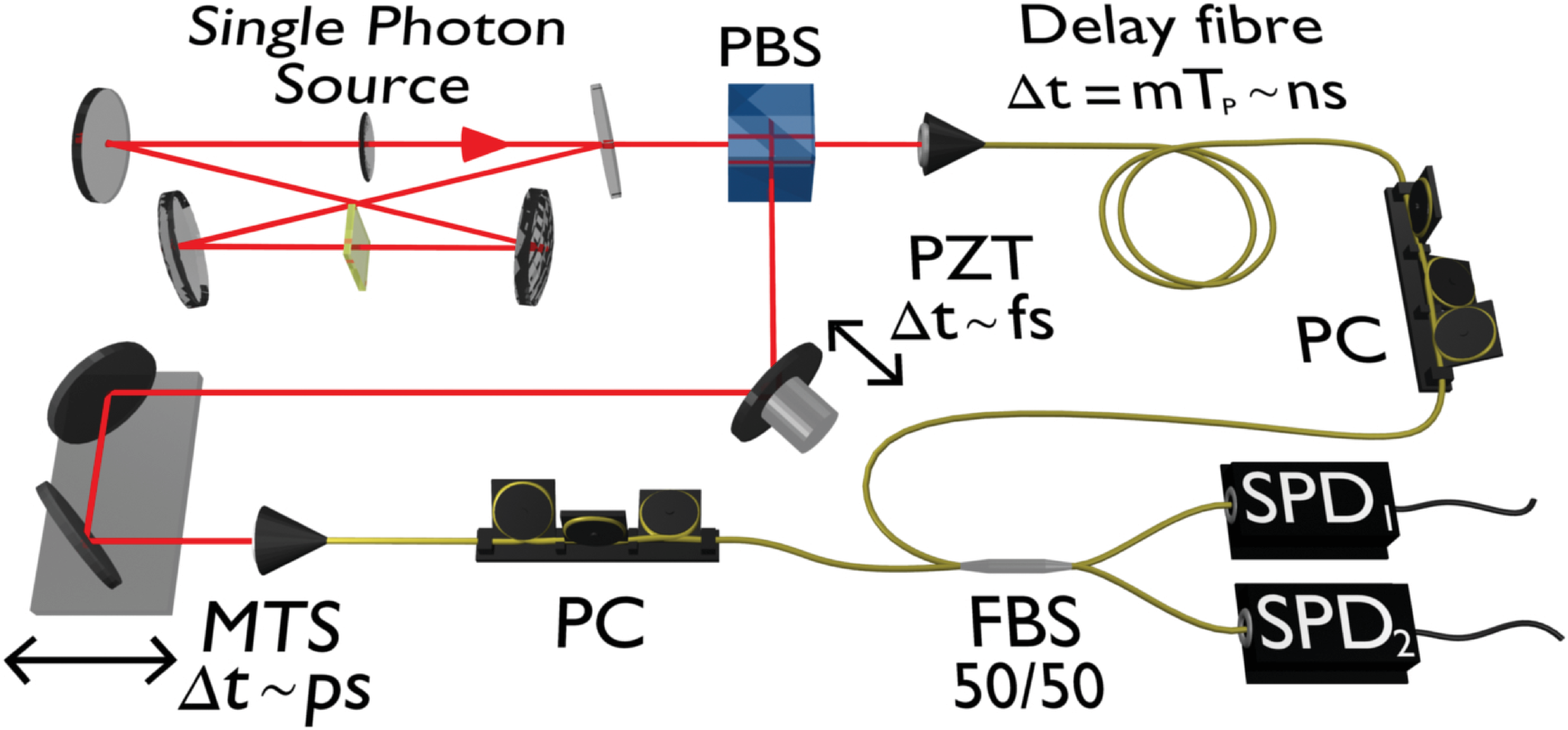}
\vspace{-7mm}
\caption{Schematic showing methods for coarse, intermediate, and fine photon delays before the 50/50 fibre beam splitter (FBS). 
PBS, polarizing beam splitter; $m \in \mathds{N}_0$; $T_p$, physical round trip time; PZT, piezoelectric transducer; PC, polarisation controller; MTS, motorized translation stage; SPD, single photon detectors. 
\emph{Coarse} (right top): Optical fibres provide delays on the scale of multiple round trips; \emph{Intermediate} (left lower corner): the motorised translation stage allows scanning over nanosecond ranges; \emph{Fine} (centre): wavelength scale resolution is achieved with the scanning piezo controller.}
\vspace{-4mm}
\label{fig:delay}
\eef

\emph{Quantum interference with a biphoton frequency comb.} --- Down conversion in our source can be described using the standard interaction Hamiltonian for SPDC of type-II~\cite{Rubin1994}. In our case this Hamiltonian leads to results differing from those found in~\cite{Herzog2008,Scholz2009b}: the flip trick changes the photons' polarization with each physical round trip. Therefore, the two photons can be detected in either: the \emph{even comb}, where the photons have orthogonal polarization and a detection time difference $\delta t$ equal to even multiples of the physical cavity round trip time ($T_p^e \equiv 2nT_p$, $n \in \mathds{N}_0$); or the \emph{odd comb}, where the photons have the same polarization and a time difference equal to odd multiples ($T_p^o \equiv (2n+1)T_p$). This results in the state,
\begin{equation}
	\label{eq:total_state}
	\ket{\psi} \propto \sqrt{2}\ket{1_{\text{H}}1_{\text{V}}}_{\delta \mathrm{t}=T_p^e} + (\ket{2_{\text{H}}0_{\text{V}}} + \ket{0_{\text{H}}2_{\text{V}}})_{\delta \mathrm{t}=T_p^o}.
\end{equation}
Note that the temporal structure eliminates overlap between the $\ket{1_{\text{H}}1_{\text{V}}}$ and $\ket{2_{\text{H}}0_{\text{V}}}$/$\ket{0_{\text{H}}2_{\text{V}}}$ terms, regardless of the photons' polarization. However, each state component in Eqn.~\eqref{eq:total_state} is in itself a frequency-entangled two-photon state containing a superposition of amplitudes for all possible even (odd) detection time differences.

The signals from detectors 1 and 2 shown in Fig.~\ref{fig:delay} are used to obtain the integrated coincidence signal, $\bar{G}^{(2)}_{1,2}(\Delta \mathrm{t})$, where $\Delta \mathrm{t}$ is the temporal delay between the photons before the beam splitter. Averaging the Glauber cross-correlation function over all possible detection times yields,
\begin{equation}
	\label{eq:homdip_result}
	\begin{split}
		\bar{G}^{(2)}_{1,2}(\Delta \mathrm{t}) &= \frac{1}{4} \big( 1 - f(2\Delta \mathrm{t}) \big) + \frac{1}{2} \sin^2 (\omega_0 \Delta \mathrm{t}),
	\end{split}
\end{equation}
where $f$ is a function of the temporal amplitude of the photons and $\omega_0$ is the central single photon frequency. The two distinct contributions in this cross-correlation function arise respectively from the even and odd frequency combs.

The first term of Eqn.~\eqref{eq:homdip_result} is due to the even comb --- the $\ket{1_{\text{H}}1_{\text{V}}}$ state --- and describes the resulting destructive two-photon interference at the 50/50 beam splitter. 
It contains the only dependence of $\bar{G}^{(2)}_{1,2}(\Delta \mathrm{t})$ on the temporal amplitude of the two-photon state, via the function
\begin{equation}
	\label{eq:dipfunction}
f(2\Delta \mathrm{t}) = \e{-2\pi \gamma \abs{\Delta \mathrm{t}}}\big( 1 + 2\pi \gamma \abs{\Delta \mathrm{t}} \big) \smashoperator[r]{\sum_{m}} h\big( 2(\Delta \mathrm{t} - m T_p) \big).
\end{equation}
Here $\gamma$ is the cavity linewidth, $h$ is the dip-shape dependent on the filters in the setup, and $m\in\mathds{Z}$. 
This shows that the photons not only interfere destructively around zero delay but also at delays which are an integer multiple of the physical round trip time $T_p$ caused by the cavity.
The shape of each separate dip is identical to the shape $h(t)$ observed in absence of the cavity function.

The second term in Eqn.~\eqref{eq:homdip_result} is the result of the odd comb --- the two-photon NOON-state~\cite{Lee2002,Dowling1998} component. Here, the delay $\Delta \mathrm{t}$ does not change the detection time difference of the two photons, but introduces a relative phase $\varphi \propto \omega_0 \Delta \mathrm{t}$. The beam splitter translates this relative phase into oscillations of the NOON state: $\varphi = 0$ leaves it unaffected while $\varphi = \pi$ transforms into the $\ket{11}$ state at the output.

In our experiment, the photons are separated deterministically after the cavity with a polarising beam splitter (PBS) before being coupled into single mode fibre.  HOM interference occurs at a $50/50$ fibre beam splitter (FBS), with polarisation control on both input ports. The two outputs are connected to silicon avalanche photon detectors and their signals recorded with a time-tagging module, Fig.~\ref{fig:delay}.

\beff
\includegraphics[width=\linewidth]{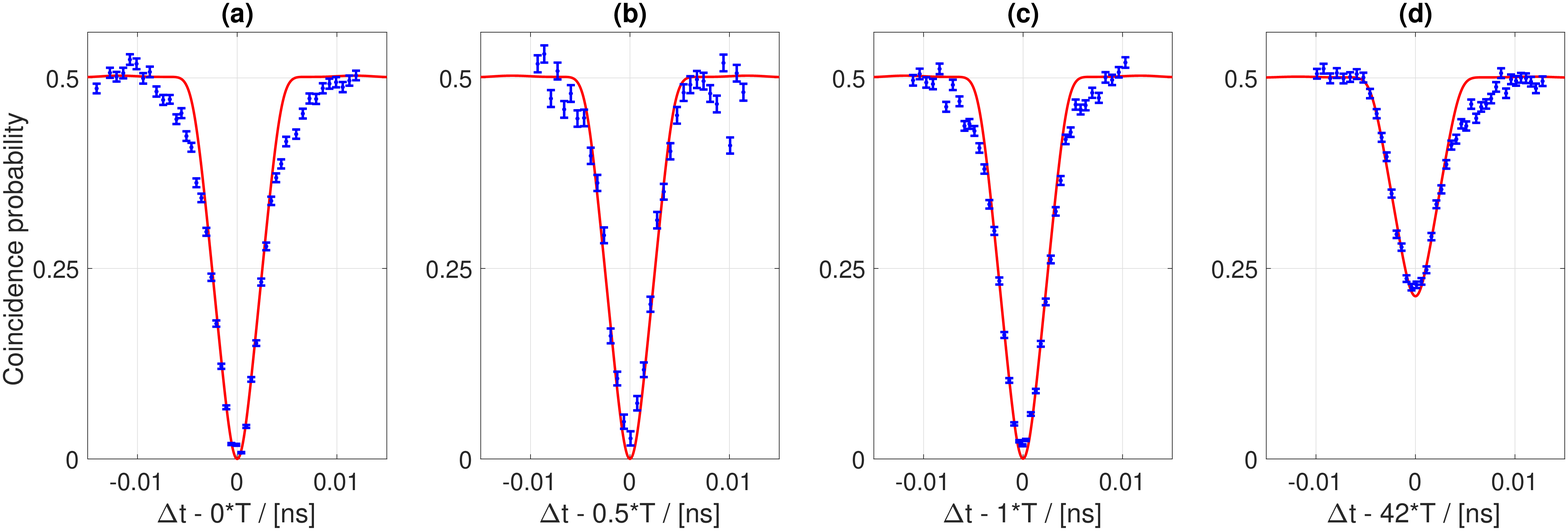}
\caption{Coincidence probability of even-comb photons in a HOM interference experiment for selected coarse time delays $\Delta \mathrm{t} \propto T (=8.28$~ns), data in blue, theory in red. 
Error bars are dominated by the uncertainty in the coincidence counts. Delays are introduced by a set of optical fibres of length 0, $\rfrac{1}{2}$, 1 and 42 round trips, corresponding to a free space path differences up to 105~m between the photons. 
Intermediate time delays between data points within (a-d) are achieved with a pair of mirrors mounted on a motorised translation stage with step sizes between 0.5--1.3~fs (150--400~$\mu$m). 
See supplementary material for additional plots.}
\vspace{-5mm}
\label{fig:HOM}
\eeff

\emph{Even-comb quantum interference.} --- Fig.~\ref{fig:HOM}\hyperref[fig:HOM]{a} shows the central HOM dip originating from the even-comb contributions in Eqn.~\eqref{eq:total_state} as the photon delay varies over a time scale of tens of picoseconds (intermediate delay).   Similar traces at positions selected with coarse delays achieved by combining fibres equivalent to $\rfrac{1}{2}$, 1, and 42 round trip times are shown in Fig.~\ref{fig:HOM}\hyperref[fig:HOM]{b-d}
(additional dips at 2, 4, and 40 round trip times and a detailed description of the procedure for collecting and post-processing of the data can be found in the supplementary material). The data points (blue) agree well with the fully-constrained theoretical model (red), where the model parameters are experimentally-determined values for linewidth, FSR, phase matching envelope and implemented narrowband filters in Eqn.~\eqref{eq:homdip_result} at constant phase. The visibility of HOM interference is given by $V {=} (P_{max}{-}P_{min})/P_{max}$, with $P_{min (max)}$ the minimal (maximal) coincidence probability. 
At zero time delay we observe near-ideal visibility, $V=(98.4\pm1.7)$\%.

\begin{figure}[!b]
\vspace{-5mm}
 \includegraphics[width=.98\linewidth,trim={0cm 0cm 0cm 0cm},clip]{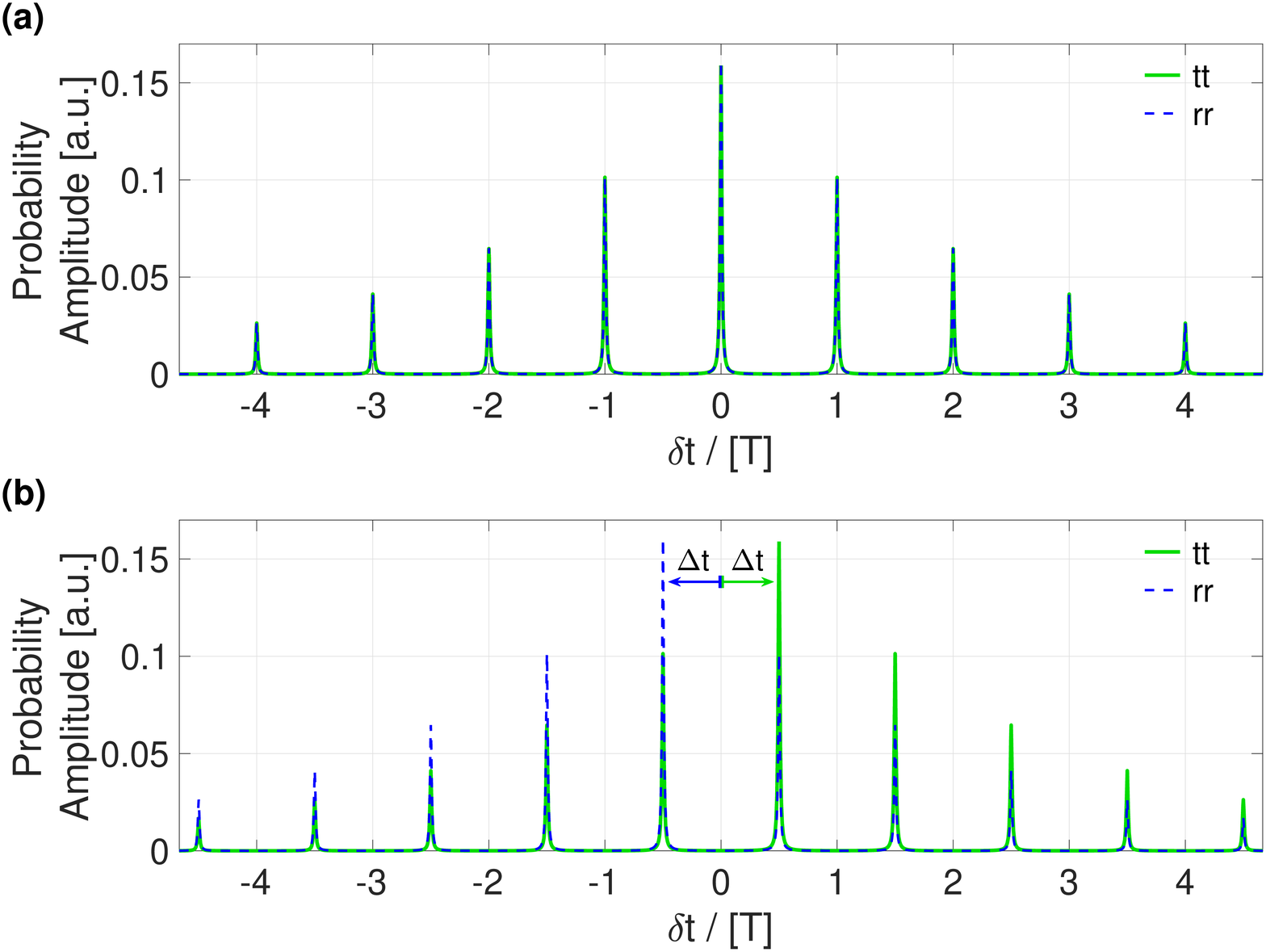}
\vspace{-1mm}
\caption{Detection probability amplitudes of the reflected-reflected (rr) and transmitted-transmitted (tt) photon paths in the HOM experiment. 
(a) No temporal delay, $\Delta \mathrm{t}=0$. 
Amplitudes cancel each other out perfectly. (b) Revivals occur when the photons are delayed by $\Delta \mathrm{t} = mT_p =m * \rfrac{T}{2}$ ($m\in\mathds{Z}$), here shown for $m=1$. Both detection probability amplitudes shift by the implemented delay but in opposite directions, overlapping the combs again.}
\label{fig:HOM_revive}
\end{figure}

\begin{figure}[b]
\vspace{-3 mm}
\includegraphics[width=\linewidth]{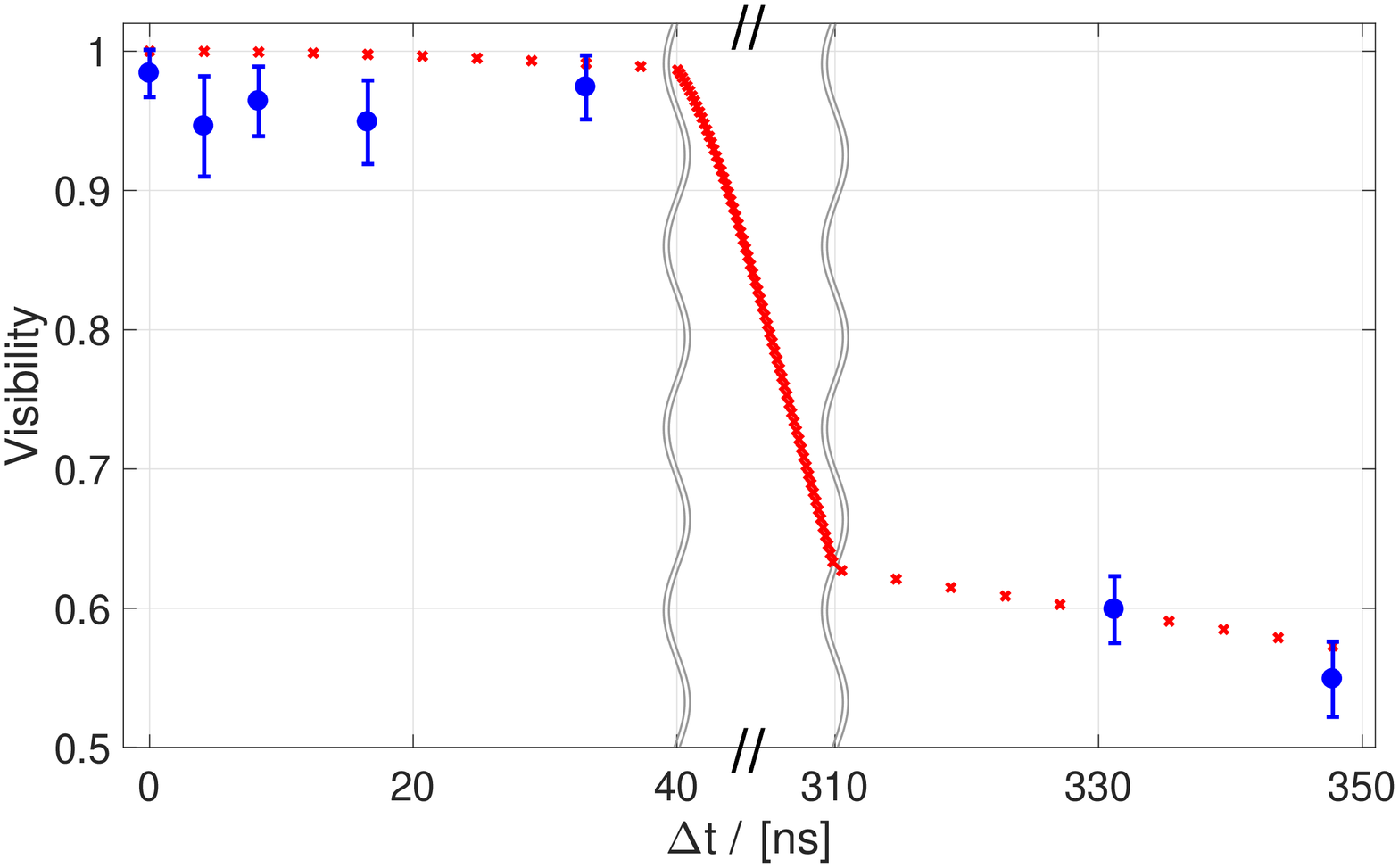}
\caption{HOM interference visibilities as a function of coarse round-trip delays. Experimental data (blue) and theoretical predictions (red) for $0, \rfrac{1}{2},1,2,4,40$, and 42 RT delays. Error bars capture the uncertainties in the photon detection and counting modules.}
\label{fig:HOM_vis}
\end{figure}

\begin{figure*}[t]
\includegraphics[width=\linewidth]{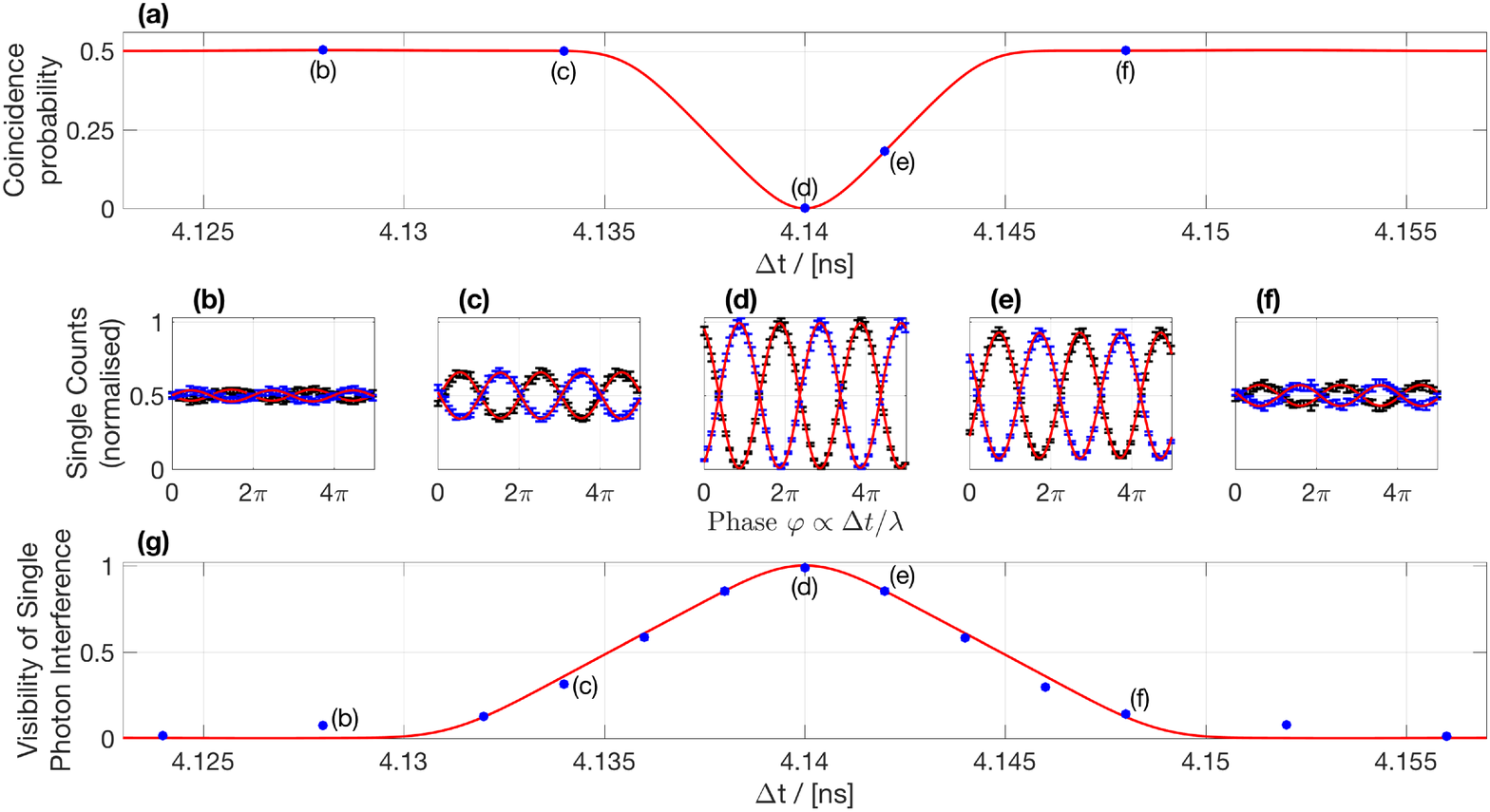}
\vspace{-5mm}
\caption{Single photon interference. 
(a) Theoretical coincidence probability between the two detectors (= HOM dip) at $\Delta \mathrm{t} = T/2$ in order to indicate the rough positions used to obtain the oscillations shown in (b-f). The delay is implemented with a set of mirrors mounted on a translation stages, see Fig.~\ref{fig:delay}. (b-f) The oscillations of the single photon counts on the two detectors in blue and black, respectively. The fit in red illustrates the expected periodicity of one wavelength and a maximal visibility for (d), at the bottom of the dip. The phase delay is realized by a mirror mounted on a piezoelectric transducer. (g) Interference visibilities, data in blue with error bars smaller than the dots, theory in red. Captions next to the data correspond to the relevant oscillation and positions on the HOM dip.}
\label{fig:SPI_vis}
\vspace{-3mm}
\end{figure*}

When single photons arrive from either side of the 50/50 beam splitter, the cases where either both photons are transmitted (tt) or reflected (rr) interfere destructively, giving rise to the HOM dip~\cite{Hong1987},  Fig.~\ref{fig:HOM}\hyperref[fig:HOM]{a}. All temporal components of the even-comb biphoton state overlap as shown in Fig.~\ref{fig:HOM_revive}\hyperref[fig:HOM_revive]{a}, and thus interfere. If a temporal delay $\Delta \mathrm{t}$ is introduced, the tt and rr detection probability amplitudes are both shifted, however, in opposite directions.  This leads to a relative shift of $2 \Delta \mathrm{t}$ so that the amplitudes re-overlap for delays matching an integer multiple of $\rfrac{T}{2}$, $\Delta \mathrm{t} = m*\rfrac{T}{2}$ with $m \in \mathds{Z}$, as illustrated in  Fig.~\ref{fig:HOM_revive}\hyperref[fig:HOM_revive]{b}, resulting in HOM dip revivals, Fig.~\ref{fig:HOM}\hyperref[fig:HOM]{b-d}. We emphasize that the somehow counter intuitive revival period is ensured by the temporal entanglement of the two photons.

The relative shift of the biphoton detection amplitudes lowers their quantitative overlap --- especially visible  around the centre of Fig.~\ref{fig:HOM_revive}\hyperref[fig:HOM_revive]{b} --- which further decreases with increasing temporal shift. This mismatch subsequently reduces the visibility of the interference as illustrated in Fig.~\ref{fig:HOM_vis} (see supplementary material for table of visibility values). The small discrepancies between the measured data and theoretical predictions in Fig.~\ref{fig:HOM_vis} arise from residual distinguishability in the polarisation degree of freedom, and from the beam splitter reflectivity varying slightly from 50\%. Observing the HOM dips with hundreds-of-nanosecond delays is proof of the long coherence time and narrow linewidth of our photons, while requiring the matching arrival times of the photons on the picosecond scale demonstrates the mode-locked state of the generated photon pairs.
 
\emph{Odd-comb quantum interference.} --- The integrated single photon count rate of each individual detector, 
\begin{equation}
	\label{eq:singlecounts}
	\begin{multlined}
		\bar{G}_{1/2}(\Delta \mathrm{t}) =\\
		1 \pm \cos (\omega_0 \Delta \mathrm{t})  \e{-\pi \gamma \abs{\Delta \mathrm{t}}} (1+\pi \gamma \abs{\Delta \mathrm{t}}) \smashoperator[lr]{\sum_{m \text{ odd}}} h\bigl(\Delta \mathrm{t} - mT_p\bigr),
	\end{multlined}
\end{equation}
oscillates with the relative delay $\Delta \mathrm{t}$. 
This results from the interference of the probability amplitudes corresponding to the $\ket{1_{\text{H}}1_{\text{V}}}$ and $\ket{2_{\text{H}}0_{\text{V}}} + \ket{0_{\text{H}}2_{\text{V}}}$ components of the state, which is only visible at delays close to odd multiples of $T_p$. To understand this, note that  originally the detection time difference of the two photons is an even (or odd) multiple of the physical cavity round-trip time for the $\ket{1_{\text{H}}1_{\text{V}}}$ (or $\ket{2_{\text{H}}0_{\text{V}}}+\ket{0_{\text{H}}2_{\text{V}}}$) state component. In principle, these contributions can be distinguished and no interference takes place. However, as the delay $\Delta \mathrm{t}$ only affects the detection time difference for the $\ket{1_{\text{H}}1_{\text{V}}}$ component, the two contributions become indistinguishable if $\Delta \mathrm{t} \approx T_p^o$, and interference is observed. The visibility of the oscillations in Eqn.~\eqref{eq:singlecounts} maximises at the centre of the half-round trip HOM dip, where the even and odd comb cannot be distinguished temporally. The interference itself is modulated across twice the width of the HOM dip as for first-order interference $\Delta \mathrm{t}$ only enters once in the relative shift of the probability amplitudes.

We emphasize that this is the result of first-order interference between the odd and even comb and consequently has to be distinguished from the oscillation of $\bar{G}^{(2)}_{1,2}(\Delta \mathrm{t})$ in Eqn.~\eqref{eq:homdip_result}, which is a result of the second-order interference of only the contributions from the NOON state. 

To record these fast nanometre-scale oscillations, we require time delays with femtosecond precision that are implemented with a mirror mounted on a piezo-electric transducer (PZT). Oscillations at selected points can be seen in Fig.~\ref{fig:SPI_vis}\hyperref[fig:SPI_vis]{b-f}, with their corresponding position on the HOM dip around $T/2$, Fig.~\ref{fig:SPI_vis}\hyperref[fig:SPI_vis]{a}. Following Eqn.~\eqref{eq:singlecounts}, the expected visibilities are compared to the measured results in Fig.~\ref{fig:SPI_vis}\hyperref[fig:SPI_vis]{g} (the plots of all oscillations are presented in the supplementary material). Comparing Fig.~\ref{fig:SPI_vis}\hyperref[fig:SPI_vis]{a} and Fig.~\ref{fig:SPI_vis}\hyperref[fig:SPI_vis]{g} clearly illustrates the expected broader range of single oscillations, with visibilities up to 0.5 outside the actual HOM dip.

\emph{Discussion.} --- We demonstrate the first non-classical interference between photons delayed by more than $\rfrac{1}{3}$ of a microsecond --- equivalent to 105~metre path length difference --- measuring a visibility of 55\% at that delay, 96\% of the theoretically predicted value. Therefore, this source can naturally be exploited in quantum networks based on time resolved correlation measurements~\cite{Tamma2015,Laibacher2017}, including multi-boson correlation sampling  schemes~\cite{Laibacher2015,Laibacher2018}. Additionally, our source achieves a heretofore unachieved spectral brightness of $(4.4~{\pm}~0.4){\times}10^{3}~\textrm{photon~pairs} / \textrm{s~mW~MHz}$. 

The source is a novel metrological tool that exhibits different kinds of quantum interference depending on which frequency comb is temporally accessed --- a quantum brush if you will. By experiencing HOM interference with phase-sensitive NOON-state super-resolution fringes at $2 \omega_0 \Delta \mathrm{t}$, and simultaneously singles oscillations at $\omega_0 \Delta \mathrm{t}$ --- with twice the spatial or temporal displacement range as the HOM interference --- our source allows enhanced precision in distance sensing of sub-wavelength features in a quantum-secured way. The HOM interference will vanish if the state of either photon is altered, allowing applications such as establishing a quantum-secured optical perimeter. Furthermore the entangled frequency combs in our source are a promising resource for frequency-multiplexed quantum information processing~\cite{Lu2018,Lukens2017}.

\emph{Acknowledgements} --- This work was supported by in part by the Australian Research Council Centres of Excellence in Engineered Quantum Systems (CE110001013, CE170100009) and Quantum Computing and Communication Technologies (CE110001027). AGW acknowledges the University of Queensland Vice-Chancellor's Research and Teaching Fellowship. VT was partially supported by the Army Research Laboratory (W911NF-17-2-0179).
SL acknowledges support by a grant from the Ministry of Science, Research and the Arts of Baden-W\"urttemberg (Az: 33-7533-30-10/19/2).

%


\clearpage
\onecolumngrid
\begin{center}
\textbf{\large Supplemental Material: Hectometer revivals of quantum interference}
\end{center}
\vspace{7mm}
\twocolumngrid

\setcounter{equation}{0}
\setcounter{figure}{0}
\setcounter{table}{0}
\setcounter{page}{1}
\makeatletter
\renewcommand{\theequation}{S\arabic{equation}}
\renewcommand{\thefigure}{S\arabic{figure}}
\renewcommand{\thetable}{S\arabic{figure}}
\renewcommand{\bibnumfmt}[1]{[S#1]}
\renewcommand{\citenumfont}[1]{S#1}

\section{The state}
The half-wave plate inside of the cavity causes the biphoton at the output of the source to be in the state
\begin{equation}
	\label{eq:supp:total_state}
	\ket{\psi} \propto \sqrt{2}\ket{1_{\text{H}}1_{\text{V}}}_{\delta t=T_p^e} + (\ket{2_{\text{H}}0_{\text{V}}} + \ket{0_{\text{H}}2_{\text{V}}})_{\delta t=T_p^o}
\end{equation}
which contains components $\ket{2_{\text{H}}0_{\text{V}}}$ and $\ket{0_{\text{H}}2_{\text{V}}}$ (the odd comb with $T_p^o \equiv (2m+1) T_p$, $m \in \mathds{Z}$) where both photons exhibit equal polarization in addition to the state $\ket{1_{\text{H}}1_{\text{V}}}$ (the even comb with $T_p^e \equiv 2mT_p$).
Using the field operators
\begin{equation}
\label{eq:supp:field_operators}
	\hat{E}_{\lambda}^{(+)}(t) \coloneqq \frac{1}{\sqrt{2\pi}}\int \d{\nu} \hat{a}_{\lambda}(\omega_0+\nu) \e{-\ii \nu t}
\end{equation}
and the notation $\cdot \conv \cdot$ for the convolution, we find the expressions in the temporal domain 
\begin{align}
	\label{eq:supp:HV}
	\ket{1_{\text{H}}1_{\text{V}}} &\propto \iint \d{t} \d{u}  \bigl(F_{\text{e}} \conv \Phi \bigr)(t-u) \hat{E}^{(-)}_{H}(t) \hat{E}^{(-)}_{V}(u) \ket{0}
	\intertext{and}
	\label{eq:supp:HH}
	\ket{2_{\text{H}}0_{\text{V}}} &\propto \iint \d{t} \d{u}  \bigl(F_{\text{o}} \conv \Phi \bigr)(t-u) \frac{\hat{E}^{(-)}_{H}(t) \hat{E}^{(-)}_{H}(u) }{\sqrt{2}} \ket{0}.
\end{align}
The fully vertically polarized component can be written analogously to Eqn.~\eqref{eq:supp:HH}.

For all three components, the temporal two-photon amplitudes can be expressed as a convolution of the two-photon amplitude $\Phi(\tau)$ for free-running SPDC with a comb-like function $F_{\text{e/o}}(\tau)$ describing the repeated reflection of the photons inside of the cavity.
This convolution is a result of the multiplication of the phase matching function with the cavity line spectrum.
More specifically, $\Phi(\tau)$ is given by the Fourier transform of the standard phase matching function (if filters are present, it is a convolution between this Fourier transform and the time response functions of the filters).

On the other hand, the functions
\begin{align}
	\label{eq:supp:even_odd_decayfunction}
	F_{\text{e}}(\tau) &\coloneqq \sum_{m=-\infty}^{\infty} \e{-\pi \gamma \abs{2m}T_p} \delta(\tau - 2mT_p)
	\intertext{and}
	F_{\text{o}}(\tau) &\coloneqq  \sum_{m=-\infty}^{\infty} \e{-\pi \gamma \abs{2m+1}T_p} \delta(\tau - (2m+1)T_p)
\end{align}
reflect the discrete spectrum of the cavity and are damped with the cavity decay rate $\gamma$.

For $\ket{1_{\text{H}}1_{\text{V}}}$ and without additional delays, the two photons can only be detected with a detection time difference $\delta \mathrm{t}$ (approximately) equal to even multiples of the physical cavity round trip time $T_p$.
If the photons exit the cavity with identical polarization however, the time difference $\delta \mathrm{t}$ is equal to an odd multiple of $T_p$.
The difference in the relative delays is understood as follows:
Since the crystal is tuned for type-II SPDC, one H- and one V-polarized photon is created in a down-conversion event.
Subsequently, the photons traverse the cavity multiple times before exiting.
At each round trip, the HWP causes the polarization of each photon to change from H to V or vice versa.
Therefore, it is only possible to detect both an H- and a V-polarized photon if the photons either both took an even number of round trips (keeping their original polarization) or both took an odd number of round trips (interchanging their original polarization).
In the other cases, one of the photons keeps its original polarization while the second one switches polarization after one round trip, resulting in an identical polarization of the photons at the output.
The resulting temporal structure, described by Eqns.~\eqref{eq:supp:HV} and \eqref{eq:supp:HH}, excludes overlap between the components $\ket{1_{\text{H}}1_{\text{V}}}$ and $\ket{2_{\text{H}}0_{\text{V}}}$/$\ket{0_{\text{H}}2_{\text{V}}}$, regardless of polarization, because the two-photon amplitudes as functions of the relative delay do not overlap.

\section{The HOM dip and HOM revivals}
The integrated coincidence signal $\bar{G}_{12}(\Delta t)$ between the two detectors 1 and 2 discussed in Eqn.~(2) of the main letter can be calculated by averaging the Glauber cross-correlation function over all possible detection times, yielding
\begin{equation}
	\label{eq:supp:cross_correlation}
	\begin{multlined}
		\bar{G}_{12}(\Delta t) \coloneqq
		\\
		\iint \d{t} \d{u} \bra{\psi} \hat{E}^{(-)}_{1}(t)\hat{E}^{(-)}_{2}(u)\hat{E}^{(+)}_{2}(u)\hat{E}^{(+)}_{1}(t) \ket{\psi}.
	\end{multlined}
\end{equation}
In order to evaluate this term, we express the field operators at the detectors in terms of the field operators $\hat{E}^{(\pm)}_{H/V}(t)$. Using the 50/50-beamsplitter transformation and taking into account that the vertically polarized photons are delayed by $\Delta t$ \big(which effectively maps $\hat{E}^{(+)}_{V}(t) \rightarrow \exp(+\ii \omega_0 \Delta t)\hat{E}^{(+)}_{V}(t-\Delta t)$ for the field operators as defined in Eqn.~\eqref{eq:supp:field_operators}\big), we arrive at
\begin{equation}
	\label{eq:supp:field_operator_connection}
	\begin{aligned}
		\hat{E}^{(+)}_1(t) &= \frac{1}{\sqrt{2}}\big( \hat{E}^{(+)}_H(t) + \e{+\ii\omega_0\Delta t} \hat{E}^{(+)}_V(t-\Delta t) \big)\\
		\hat{E}^{(+)}_2(t) &= \frac{1}{\sqrt{2}}\big( \hat{E}^{(+)}_H(t) - \e{+\ii\omega_0\Delta t} \hat{E}^{(+)}_V(t-\Delta t) \big).
	\end{aligned}
\end{equation}
After inserting these relations together with the state of Eqn.~\eqref{eq:supp:total_state} into Eqn.~\eqref{eq:supp:cross_correlation}, a straightforward calculation yields the final result
\begin{equation}
	\label{eq:supp:homdip_result}
	\begin{split}
		\bar{G}_{12}(\Delta t) &= \frac{1}{4} \big( 1 - f_{ee}(2\Delta t)/f_{ee}(0)  \big) + \frac{1}{2} \sin^2 (\omega_0 \Delta t)
	\end{split}
\end{equation}
This cross-correlation function contains two distinct contributions which arise due to the even-comb ($\ket{1_{\text{H}}1_{\text{V}}}$) or odd-comb ($\ket{2_{\text{H}}0_{\text{V}}} + \ket{0_{\text{H}}2_{\text{V}}}$) component of the input state.
Due to the symmetries at the beam splitter, the terms describing interference between the two components drop out and the remaining terms can be attributed to either the state component with different photon polarizations or to the one with identical photon polarizations.
The first term is associated to the $\ket{1_{\text{H}}1_{\text{V}}}$ state component and describes the resulting destructive two-photon interference at a 50/50-beam splitter.
It contains the only dependence of $\bar{G}_{12}(\Delta t)$ on the temporal amplitudes $ \big( F_{\text{e/o}}\conv \Phi \big)(t-u)$ of the states via the function
\begin{multline}
	\label{eq:supp:dipfunction}
		f_{ee}(2\Delta t)
		\\
		\coloneqq \iint \d{t} \d{u} \big( F_{\text{e}} \conv \Phi \big)^*(t-u+2\Delta t) \big( F_{\text{e}} \conv \Phi \big)(t-u) \\
				     \propto \e{-2\pi \gamma \abs{\Delta t}}\big( 1 + 2\pi \gamma \abs{\Delta t} \big) \smashoperator[r]{\sum_{m}} h\big( 2(\Delta t - T_p^e) \big)
\end{multline}
describing the destructive two-photon interference between the photons in the state $\ket{1_{\text{H}}1_{\text{V}}}$ before the delay $\Delta \mathrm{t}$ in front of the symmetric beam splitter.
Due to the special temporal structure of the biphoton stemming from the discrete mode structure of the cavity, the photons do not only interfere destructively around zero delay but also at delays which are an integer multiple of the physical cavity round trip time $T_p = T/2$.
Nevertheless, due to the damping of the cavity, the visibility of these dips decreases as the difference between the H and V paths increases.
The shape of the each separate dip is identical to the shape $h(t)$ observed in absence of the cavity function.
If no filters are taken into account, it is given by a triangular function
\begin{equation}
	\label{eq:supp:triangulardip}
	h(t) =
	\begin{dcases}
		1 - \frac{\pi \delta\nu \abs{t}}{2.783} & \abs{t} \leq 2.783/(\pi\delta\nu) \\
		0 & \text{else}.
	\end{dcases}
\end{equation}
Here, $\delta\nu$ is the phase matching bandwidth defined at FWHM of the $\operatorname{sinc}^2$ phase matching function.
The effect of filters in the setup can be fully described as a modification of $h(t)$, mainly leading to a broadening of the dip.

While the first term in the cross-correlation Eqn.~\eqref{eq:supp:homdip_result} can be fully attributed to the $\ket{1_{\text{H}}1_{\text{V}}}$ state component, the second term is the result of the equal polarization component $\ket{2_{\text{H}}0_{\text{V}}} + \ket{0_{\text{H}}2_{\text{V}}}$ of the state.
The oscillatory behavior of the latter is easily understood by noting that this state component is effectively a two-photon NOON-state.
While the delay of the V-polarized photons does not change the relative detection times of the two photons for this state component, it does result in a relative phase $\varphi \propto \omega_0 \Delta \mathrm{t}$.
A symmetric beamsplitter then translates this phase into different outputs: for $\varphi = 0$ the state remains unchanged while for $\varphi = \pi$ the output is now the state $\ket{11}$.

It is worth noting that the NOON-state contribution in Eqn.~\eqref{eq:supp:homdip_result} can be eliminated around delays $\Delta t=2kT_p$, $k \in \mathds{Z}$. 
For these delays, photons created in the $\ket{1_{\text{H}}1_{\text{V}}}$ state are still detected with a time difference which is an even multiple of the physical cavity round trip time $T_p$, while the photons created in the $\ket{2_{\text{H}}0_{\text{V}}} + \ket{0_{\text{H}}2_{\text{V}}}$ arrive with a time difference which is an odd multiple of $T_p$.
Therefore, filtering for time-resolved coincidence events where the detection times differ by (approximately) $2kT_p$, the oscillatory behavior of Eqn.~\eqref{eq:supp:homdip_result} can be suppressed.

All measured HOM revivals - specifically at delays of $\rfrac{1}{2}$, 1, 2, 4, 40 and 42 round trips - are presented in Fig.~\ref{fig:HOM_all}. 
These were obtained following the procedures detailed in the next section. Fig.~\ref{fig:HOM_all}(\hyperref[fig:HOM_all]{a,b,g}) are presented in Fig. 4 of the main paper.

The numeric values of the theoretical and measured HOM dip visibilities corresponding to the measurements in Fig.~\ref{fig:HOM_all} can be found in Tab.~\ref{tab:dip_vis}. The relative visibilities of each measurement have been included as a reference for how close the experimental values are to the expected ones --- specifically $V_{rel} = V_{exp}/V_{th}$. All visibilities are stated as percentages.

\beff
\includegraphics[width=\linewidth]{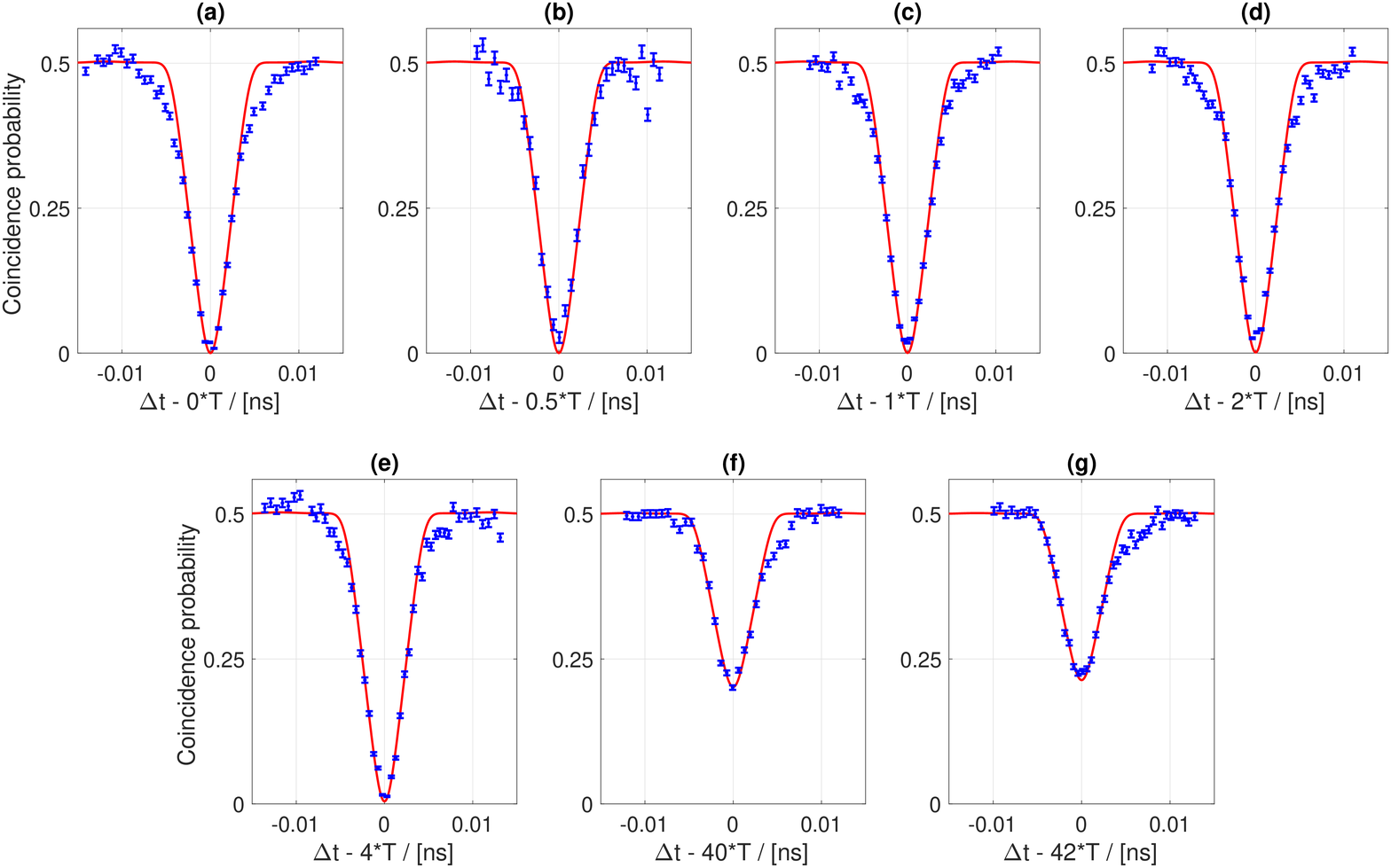}
\caption{Coincidence probability in a HOM interference experiment for various coarse time delays $\Delta \mathrm{t} \propto T (=8.28$~ns), data in blue, theory in red. Error bars are dominated by the uncertainty in the coincidence counts. Different delays are introduced by a set of optical fibres of length 0, $\rfrac{1}{2}$, 1, 2, 4, 40 and 42 round trips, corresponding to free-space path differences up to 105~m between the photons. The finer time delays between data points within (a-g) are achieved with a pair of mirrors mounted on a motorised translation stage with a step size between $0.5 - 1.3$~fs ($150 - 400$~$\mu$m).}
\label{fig:HOM_all}
\eeff

\begin{table}[!h]
\begin{center}
\begin{tabular}{|c|c|c|c|}
		\hline
		Delay $/$ cavity & Visibility $V_{th}$ & Visibility $V_{exp}$ & Visibility $V_{rel}$\\
        round trip time & (theoretical) & (experimental) & (relative)\\
		\hline \hline
		0 & 100 & $98.4 \pm 1.7$ & $98.4 \pm 1.7$\\
        $\rfrac{1}{2}$ & 99.99 & $95 \pm 4$ & $95 \pm 4$\\
		1 & 99.94 & $96 \pm 3$ & $96 \pm 3$\\
		2 & 99.77 & $95 \pm 3$ & $95 \pm 3$\\
		4 & 99.12 & $97 \pm 2$ & $98 \pm 2$\\
		40 & 59.7 & $60 \pm 2$ & $101 \pm 3$\\
		42 & 57.3 & $55 \pm 3$ & $96 \pm 5$\\
		\hline        
\end{tabular}
\caption{HOM visibilities for different delays. Uncertainties dominated by counting errors. 
Possible distinguishability of the photons before the beam splitter, e.g. in polarisation, or imperfections of the 50/50 beam splitter are not included.}
\vspace{-5mm}
\label{tab:dip_vis}
\end{center}
\end{table}

\vspace{0.2cm}
\section{Data Collection and Post-processing Procedure}
Unlike free-running SPDC sources, cavity-enhanced sources experience drifts in the photon pair production rate based on slight shifts in the resonance condition. 
To compensate for this effect, we ensure each data point for the HOM dips consists of a total number of $(3~\textrm{or}~4)\times 10^5$~counts per detector by varying the collection time window between $7-10$~seconds at a single photon rate $\sim 40$~kHz. 
The cross-correlation function $\bar{G}_{12}(\Delta t)$ is measured by recording the time-resolved coincidences $\bar{G}_{12}(\delta t, \Delta t)$ between the detectors and subsequently summing over all $\delta\mathrm{t}$. 
As our source produces many frequency modes, which are mode-locked due to the cavity structure, arrival time differences $\delta t$ of the photons in the $\bar{G}_{12}(\delta t, \Delta t)$ measurement exhibit a comb-like structure spaced by the physical round trip time of the cavity~\cite{Rambach2016}. 
In order to select photon pairs arriving from opposite input ports of the beam splitter, we only consider coincidence peaks of $\bar{G}_{12}(\delta t, \Delta t)$ for detection time differences $\delta t = mT = 2mT_p$. 
This is ensured by first selecting detection events within a large $\pm 500$~ns delay window around the central peak and then post-filtering a $1.07$~ns window over individual peaks (see~\cite{Rambach2017} for more details). 
Hence, we use an effective detection window of $\sim 129$~ns. 
As mentioned, this post-selection suppresses oscillatory behavior of the NOON-state contribution in Eqn.~\eqref{eq:supp:homdip_result}.
Artificial differences between the photon count rates due to misalignment of the optics are eliminated by normalising and rescaling the coincidence data to $0.5$ outside the dip. 
All data points are background corrected and the error bars correspond to the square root of the coincidence counts. 

In order to take a single HOM interference trace, the above data acquisition process is repeated at different time delays $\Delta \mathrm{t}$, implemented solely by a motorised translation stage in one of the collection arms. 
This allows us to measure the HOM dip for approximately equal photon arrival times on the beam splitter, or $\Delta \mathrm{t} = 0$~ns, as shown in Fig.~\ref{fig:HOM_all}\hyperref[fig:HOM_all]{a}. The same procedure can be repeated with the rough fibre delays to observe revived HOM dips, as presented in the previous section.  

There are some deviations from the above mentioned data-collection and post-processing procedure for the HOM dip at $\rfrac{1}{2}$ round trip delay.
In addition to ensuring sufficient statistics are collected at an approximately constant rate, the piezo-mounted mirror was also used to keep the phase of the NOON-state contribution constant at the $\ket{11}$ output. 
This was to keep the single detections balanced between the detectors, thus mechanically suppressing the singles oscillations.
In this case, the HV pairs arrive with the same time differences at the detectors as the HH or VV photon pairs which have one physical round trip time difference. 
Therefore the post-processing procedure for such datasets differs from that of the integer round trip delays: we select coincidence peaks centred at an offset of half a round trip, subtract the predicted contributions to the coincidence counts due to HH and VV pairs and rescale the data to 0.5.

\section{Single Photon Interference}
The integrated rate
\begin{equation}
	\label{eq:supp:singlecountsdef}
	\bar{G}_{i}(\Delta t) \coloneqq \int \d{t} \bra{\psi} \hat{E}^{(-)}_{i}(t)\hat{E}^{(+)}_{i}(t) \ket{\psi}
\end{equation}
of each individual detector can be calculated via the respective first-order Glauber correlation function to determine the effect of the temporal delay of the vertically polarized photons in front of the beam splitter.
In analogy to the definition of $f_{ee}(\Delta t)$ in Eqn.~\eqref{eq:supp:dipfunction}, let us define the functions
\begin{multline}
	\label{eq:supp:functioneo}
		f_{eo}(\Delta t) \coloneqq
		\\
		\iint \d{t} \d{u} \big( F_{\text{e}} \conv \Phi \big)^*(t-u+\Delta t) \big( F_{\text{o}} \conv \Phi \big)(t-u)
\end{multline}
and
\begin{multline}
	\label{eq:supp:functionoo}
		f_{oo}(\Delta t) \coloneqq
		\\
		\iint \d{t} \d{u} \big( F_{\text{o}} \conv \Phi \big)^*(t-u+\Delta t) \big( F_{\text{o}} \conv \Phi \big)(t-u).
\end{multline}
With their help, we find after inserting the field operators from Eqn.~\eqref{eq:supp:field_operator_connection} and the state Eqn.~\eqref{eq:supp:total_state} that the rates take the form
\begin{equation}
	\label{eq:supp:singlecounts}
		\bar{G}_{i}(\Delta t) =1 \pm \Re \bigl[ \e{\ii\omega_0 \Delta t} f_{eo}(\Delta t)\bigr]/\sqrt{f_{ee}(0) f_{oo}(0)}.
\end{equation}
After evaluation of Eqns.~\eqref{eq:supp:functioneo} and \eqref{eq:supp:functionoo} and using the result from Eqn.~\eqref{eq:supp:dipfunction}, this can be further simplified to yield the result
\begin{multline}
		\bar{G}_{i}(\Delta t) = \\
		1 \pm \cos (\omega_0 \Delta t)  \e{-\pi \gamma \abs{\Delta t}} (1+\pi \gamma \abs{\Delta t}) \smashoperator[lr]{\sum_{m \text{ odd}}} h\bigl(\Delta t - mT_p\bigr)
\end{multline}
from Eqn.~(4) in the main letter.

The obvious oscillation of the rates with the relative delay $\Delta t$ is the result of the interference between the probability amplitudes corresponding to the $\ket{1_{\text{H}}1_{\text{V}}}$ and $\ket{2_{\text{H}}0_{\text{V}}}+\ket{0_{\text{H}}2_{\text{V}}}$ components of the state respectively.
We further note that the interference is only visible at delays which are close to odd multiples of $T_p$.
This can be understood if we note that, originally, the relative delay between the two photons produced by the source is an even multiple of $T_p$ in the first state component and an odd multiple in the second.
Thus, without the additional delay, it is possible to distinguish between the states of the biphoton in terms of the detection times of the photons and consequently no interference between these possibilities occurs.
This changes if a non-zero delay $\Delta t$ is introduced.
While the relative delay of the photons is not affected by $\Delta t$ if they are produced with equal polarization (either no photon is delayed or both are equally delayed), it changes if they are produced with different polarization.
If $\Delta t$ is close to an odd multiple of $T_p$, the relative delay is (approximately) equal for all possible contributions and interference can be observed.
We emphasize that this interference phenomenon is in contrast to the oscillation of $\bar{G}_{12}(\Delta t)$ in Eqn.~\eqref{eq:supp:homdip_result}, which is a result of the interference of the contributions from $\ket{2_{\text{H}}0_{\text{V}}}$ and $\ket{0_{\text{H}}2_{\text{V}}}$ only.

\beff
\includegraphics[width=\linewidth]{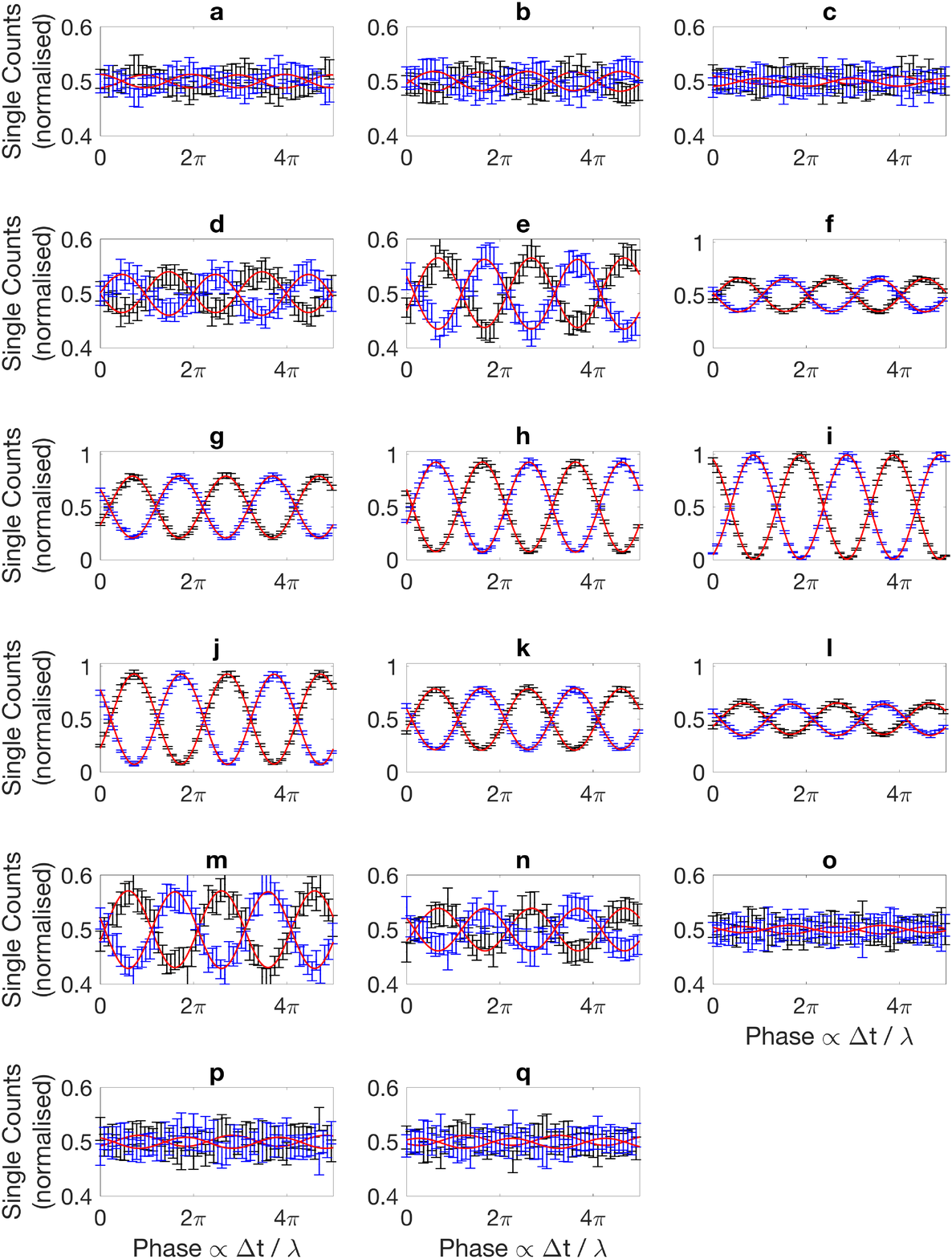}
\vspace{-25mm}
\caption{Single photon oscillations observed at different time delays. 
The intermediate time delay between measurements (a-e) and (m-q) is 4~ps, whereas measurements (f-l) are separated by 2~ps. 
The visibilities of the oscillations maximise at the point of full HOM interference (measurement (i)) and decrease around that time delay. 
The y-axis of measurements (a-e) and (m-q) have been scaled for clarity.}
\label{fig:SPI_singles_all}
\eeff

All measured single photon oscillations at half round trip time delay are presented in Fig.~\ref{fig:SPI_singles_all} in order of increasing time delay. 
These single photon count rates occur due to the even- and odd-comb amplitudes interfering to create oscillations with a period of the wavelength. 
The oscillations are visible over twice the width of the HOM dip. 
The central five measurements, namely Fig.~\ref{fig:SPI_singles_all}(\hyperref[fig:SPI_singles_all]{g-k}), are within the non-classical interference region and all had reduced $\chi^2$ values between 0.91 and 1.26.
The first and last four measurements, corresponding to Fig.~\ref{fig:SPI_singles_all}(\hyperref[fig:SPI_singles_all]{a-d,n-q}), exhibit a reduced $\chi^2$ value below 0.45, while the measurements closer to the HOM interference region, specifically Fig.~\ref{fig:SPI_singles_all}(\hyperref[fig:SPI_singles_all]{e-m}), all had reduced $\chi^2$ values between 0.58 and 1.26. Each measurement is obtained by controlling time delays with femtosecond precision through a piezo-mounted mirror, with the temporal delay between each measurement adjusted by the motorised translation stage. 
The y-axis of the first and last five measurements are scaled for clarity as the oscillations are significantly smaller outside the HOM interference region. 
Fig.~\ref{fig:SPI_singles_all}(\hyperref[fig:SPI_singles_all]{d,f,i,j,m}) are presented in Fig.~4 of the main paper. 

\end{document}